\documentclass[proceedings]{JHEP} 

\newcommand{\be}{\begin{equation}}
\newcommand{\ee}{\end{equation}}
\newcommand{\bea}{\begin{eqnarray}}
\newcommand{\eea}{\end{eqnarray}}

\newcommand{\no}{\nonumber\\}
\newcommand{\tAtwo}{\tilde{\mathbf{A}}_2}
\newcommand{\tAp}{\tilde{\mathbf{A}}_{\mbox p}}
\newcommand{\bM}{\mathbf{M}}
\newcommand{\bO}{\mathbf{O}}
\newcommand{\cS}{{\cal S}}
\newcommand{\cC}{{\cal C}}
\newcommand{\cM}{{\cal M}}

\newcommand{\rmga}{(r-\gamma_{\alpha})}

\newcommand{\NC}{N_c}

\newcommand{\CTT}{\widetilde{C}_2}

\conference{Trieste Meeting of the TMR Network on Physics beyond the SM}

\title{N=1 Super-Yang-Mills on the Lattice: Weak and Strong Coupling Limits}

\author{Emidio Gabrielli\thanks{Works made in collaboration with  
A. Donini and B. Gavela for {\it Quenched Supersymmetry} 
and A. Gonz\'alez-Arroyo
and C. Pena for {\it Strong Coupling Limit}}\\
        Address: {\rm Departamento de F\'{\i}sica Te\'orica,
        Universidad Aut\'onoma, 28049 Madrid, Spain}\\
        E-mail: \email{emidio.gabrielli@cern.ch}}

\abstract{We present a general review about the N=1 supersymmetric
$SU(\NC)$ Yang-Mills 
on the lattice focusing our attention on the quenched 
approximation in supersymmetry. 
Finally we analyse and discuss 
the recent results obtained at strong coupling
and large $N_c$ for the mesonic and fermionic propagators and 
spectrum.}

\begin{document} 

\section{Introduction}
In this paper we consider the non-perturbative aspects
of the strongly interacting supersymmetric gauge theories. \cite{npSUSY}.
In particular we concentrate our attention on the pure N=1 
Supersymmetric Yang-Mills (SYM) theory.

The fundamental question of the supersymmetry (SUSY) 
breaking of the N=1 SYM was addressed
in Refs.\cite{Windex}, \cite{VY}.
According to the general argument of the Witten index \cite{Windex}
or the 
Veneziano-Yankielowicz (VY) low energy effective theory \cite{VY}
one can conclude that:
\begin{itemize}
\item the spontaneous breaking of chiral symmetry occurs:
the gluino condensate $\langle \bar{\lambda} \lambda \rangle \ne 0$
\item 
the low energy supermultiplet is 
given by the scalar, pseudoscalar and fermion colourless bound states.
\item SUSY is not broken
\item no goldstone boson (or pion) associated with the chiral symmetry
breaking is present, as the latter is broken by the anomaly. 
\end{itemize}

A primary tool for a direct study of strongly coupled field theories
is the space-time lattice regularization.
In QCD the non perturbative phenomena such as confinement,
chiral symmetry breaking and spectra can be studied on the lattice 
by using numerical Monte Carlo simulation in the weak coupling limit. 
Nevertheless, when the supersymmetric gauge theories are formulated 
on the lattice, the following problems arise
\begin{itemize}
\item The lattice breaks the Poincar\'e group and so there are no continuum 
SUSY algebra transformations.
\item The naive lattice formulation 
breaks the balance $n^0$ {\it bosons}=$n^{0}$ {\it fermions} due to
the extra poles in the gluino propagator.
The Wilson term cures this problem, but generates a
bare mass term for the gluinos
which breaks explicitely both the chiral symmetry and supersymmetry.
\end{itemize}
The basic idea to circumvent these problems 
was proposed some time ago by Curci and Veneziano \cite{cv}.
They suggested to leave that the lattice regularization spoil SUSY and 
chiral symmetries. Then the Ward Identities (WI) in the continuum limit 
should be recovered by using appropriate renormalized operators for the
SUSY and chiral currents
\cite{cv}, such as the case of chiral symmetry in QCD \cite{boch}.
The main result is that in the continuum limit 
the chiral limit defines the SUSY point and viceversa \cite{cv}.

Recently two different collaborations \cite{montvay}-\cite{munster}
studied non perturbatively
on the lattice the spectrum of the N=1 SYM  theory following
the guide lines of Ref. \cite{cv}. In Ref. \cite{andrea}, because of the
limitation deriving from the use of computing resources, the quenched
approximation was used to study the spectrum. This approximation is
implemented by neglecting the internal gluino loops or in other words by
setting to unity the fermion determinant in the correlation functions
of composite operators. 
In the SYM theory the quenched approximation badly breaks SUSY and it cannot be
a good approximation on the basis of large $N_c$ dominance since
gluinos are in the adjoint representation of the colour group.
However the quenched numerical results for the low energy spectrum
show no deviations from the supersymmetry expectation 
within the statistical errors \cite{andrea}.

In connection with this result, in Ref. \cite{qsusy} a qualitative and
quantitative understanding of the effects of quenching in N=1 SYM theory
has been analysed in the framework of low energy effective theory.
The result is that the splitting in the mass spectrum of the low energy 
supermultiplet is connected to the changing of the anomalies 
structure induced by quenching.

Recently it has been analysed the strong coupling limit of the N=1 SYM 
on the lattice in the large $N_c$ limit
\cite{our}.
The method used in Ref. \cite{our} is based on the hopping parameter ($k$) 
expansion in 
terms of random walks which have been resummed for any value of the Wilson
parameter ($r$) in the small hopping parameter region.
Analytical results have been obtained for the propagator and
spectra of the mesonic 2-gluino  and fermionic 3-gluino operators
in terms of $r$ and $k$.
Moreover the critical lines in $k$ and $r$ space, 
where the chiral symmetry and supersymmertry can be recovered
in the continuum limit, have been
analysed for any dimension \cite{our}.

The paper is organized as follows.
In the next section we discuss the weak coupling limit 
on the lattice and summarize the approach of Ref. \cite{cv}.
In section 3 we present the results on the SUSY spectrum induced by the
quenched approximation by means of a low energy effective lagrangian.
In section 4 we discuss the strong coupling limit at large $N_c$  
and give the main results for the correlation functions 
and spectra for the 2-gluino and 3-gluino operators.
Finally in the last section we summarize our conclusions.
\section{Weak coupling limit}
The lattice chiral WI can be obtained by applying 
the chiral transformations to the $N=1$ SYM lattice. The result is given by
\footnote{the expression for the chiral currents $A_{\mu}$ and 
the pseudoscalar density $P$, together with
the N=1 Super-Yang-Mills action on the lattice, can be found in 
Refs. \cite{cv}, \cite{andrea}}
\be
\nabla_{\mu} A_{\mu} = 2 m_0 P + X_A 
\label{barechiralWI}
\ee
where $m_0$ is the gluino bare mass.
The operator $X_A$ comes from the chiral 
symmetry-breaking due the lattice spacing and it vanishes in the
continuum limit ($a\to 0$) since it is of order $O(a)$.
Nevertheless when we take the matrix elements of Eq.(\ref{barechiralWI})
between external states, the contribution of the operator $X_A$ could not 
vanish in the continuum limit. Indeed $X_A$ can induce 
divergencies of order $O(1/a)$ that compensate the explicit factor
$a$ in $X_A$ and spoil the WI in the continuum limit.

However it is possible to define a renormalized operator
$\hat{X}_{A}$ whose matrix elements are still of order $O(a)$
\cite{boch},\cite{cv}.
Due to the symmetries of the action, $\hat{X}_A$ 
can mix only with the following operators 
\bea
\hat{X}_A &=& X_A + ( Z_A - 1 ) \nabla_\mu A_{\mu} - 
{\tilde Z}_A \nabla_{\mu} A_{\mu} \no
&-& Z_Q P_{\mu\nu}{\tilde P_{\mu\nu}}+ 2 {\bar m} P
\label{xa}
\eea
where $P_{\mu\nu}$ is the lattice transcription of the field strength 
$F_{\mu\nu}$ and $\tilde{P}_{\mu\nu}$ is the dual.
Finally, by inserting Eq.(\ref{xa}) inside Eq.(\ref{barechiralWI}), we
obtain the renormalized chiral WI 
which has the good continuum limit
\cite{cv}
\be
 \nabla_{\mu} {\hat A}_{\mu} = 2 (m_0-{\bar m}) Z^{-1}_P {\hat P} + {\hat Q} 
+ O(a)
\label{renchiralWI}
\ee
provided that 
\bea
\hat A_{\mu} &=& Z_A(g_0) A_{\mu}, ~~~~\hat P = Z_P(g_0) P \no
\hat Q &=& Z_Q(g_0) P_{\mu\nu}{\tilde P_{\mu\nu}}+
{\tilde Z}_A(g_0) \nabla_{\mu}
A_{\mu} \nonumber
\eea
where the $\hat{Q}$ term reproduces the usual chiral anomaly.
It is important to note that the Eq.(\ref{renchiralWI}) 
has the same form as the continuum one provided that
we identify on the lattice the renormalized gluino mass 
$\hat{m}_{\lambda}$ as follows \cite{cv}
\be
\hat m_{\lambda} =  (m_0 - \bar{m}) Z^{-1}_P.
\label{chlim}
\ee
Finally the chiral symmetry on the lattice is recovered by tuning $m_0$ to
a critical value $m_0^{crit}$
\be
m_0^{crit} - \bar{m}_{\lambda}(m_0^{crit},g_0,r)=0
\ee
where the $\bar{m}$ term depends in general on $m_0$, $g_0$ and $r$.
Note that  $Z_A$ is of order $Z_A=1+O(g_0)$ and in the
continuum limit ($g_0\to 0$)
we obtain $Z_A=1$, as expected by the current non renormalization theorem.

In the case of SUSY WI we have an analogous result of 
Eq.(\ref{barechiralWI})
\[
\nabla_{\mu} S_{\mu} = 2 m_0 \chi + X_S\nonumber
\]
where now $S_\mu(x)$ is the bare SUSY current on the lattice
and $\chi= 1/2 P^a_{\mu\nu}\sigma_{\mu\nu}\lambda^a$
where the $\lambda^{a}$ is the gluino field. 
The expressions for $X_S$ and $S_{\mu}$ can be found in Ref.\cite{cv}.
The operator $X_S$, which is of order $O(a)$,
spoil the continuum SUSY WI, like $X_A$ in the case of chiral symmetry.\\
By applying the same method used for the chiral WI one obtains the
following renormalized SUSY WI \cite{cv}
\[
\nabla_{\mu} {\hat S}_{\mu} = 2 (m_0 -\bar{m}) Z^{-1}_{\chi} \hat \chi + 
O(a)\nonumber
\]
where
\bea
\hat \chi &=& Z_\chi \chi,~~~~
\hat S_{\mu} = Z_S S_{\mu} + Z_T T_{\mu}\no
T_{\mu}(x)&\equiv& \gamma_{\nu} P^a_{\nu\mu}(x) \lambda^a(x)
\nonumber
\eea
This result coincides with the corresponding one in the continuum, 
provided that the renormalized gluino mass $\hat m_{\lambda}$ 
is identified with
\be
\hat m_{\lambda} =  (m_0 - \bar{m}) Z^{-1}_{\chi}\nonumber
\label{gluinomass}
\ee
Then the relevant conclusion is that in the continuum limit
the chiral limit of Eq.(\ref{chlim}) defines the SUSY point  and
viceversa \cite{cv}.

The present numerical analysis implement these guidelines
for studing the spectrum of the N=1 SYM with SU(2) gauge group.
According to Veneziano-Yankielowicz \cite{VY}, 
the low-energy SUSY supermultiplet is given by the
following colourless composite fields 
\bea
S(x)&=&\bar{\lambda}^{a}(x) \lambda^{a}(x),~~~~~
P(x)=\bar{\lambda}^{a}(x)\gamma_5  \lambda^{a}(x),\no
\chi(x)&=&G^{a}_{\mu\nu}(x)\sigma_{\mu\nu}\lambda^{a}(x)
\label{VImultiplet}
\eea
where the sum on the colours is assumed.
As usual the masses are extracted from the large Euclidean-time behaviour of 
the lattice correlation functions for the corresponding operators in 
Eq.(\ref{VImultiplet}).

In Ref. \cite{andrea} the quenched approximation is used in which
dynamical gluino loops are neglected, or the fermion 
determinant $\det(K)$ is setted to 1 in the correlation functions.\footnote{
Really in the present case one has the Pfaffian instead of
$\det(K)$ since the gluinos are Majorana fields.}
In QCD the quenched approximation  
is a good one because the $\det(K)=O(1/N_c)$ in the large $N_c$ limit.
In the present case the gluinos
are in the adjoint representation of the colour group (like the gluons)
and by using naive arguments based on perturbation theory
one should expect that the quenched approximation badly breaks SUSY.
\footnote{These arguments, based on perturbation theory, do not
apply in the strong coupling limit (see section 4), where indeed
this approximation is exact at large $\NC$} 
Nevertheless the quenched results of Ref. \cite{andrea} 
show a dynamical chiral symmetry breaking and a quite degenerate spectrum
in low energy supermultiplet.
Moreover in Ref.\cite{andrea} the OZI approximation has been used.
In this approximation the diagrams which contribute to the chiral anomaly 
are neglected and
by using general arguments \cite{cv} one should expect a 
massless pseudo-goldstone boson or pion in the spectrum.

In the next section we will show how to implement the quenching 
in the fundamental theory. Then we will give an
estimation of the systematic error induced by the quenching
on the spectrum, by means of a low energy lagrangian 
approach.

\section{Quenched Supersymmetry}
In the continuum theory the on-shell action of the N=1 SYM theory is given by
\be
\label{symaction}
S_{SYM} = \int d^4 x \left \{ - \frac{1}{4} F^a_{\mu\nu} F^{a\mu\nu}
 + \frac{i}{2} \bar \lambda^a \gamma^\mu D^{ab}_\mu \lambda^b \right \} 
\label{SYM}
\ee
where $D^a_{\mu}$ is the covariant derivative acting on the gluino field $\lambda^a$.
At the classical level this action is $U(1)_A$ invariant, as well as 
scale invariant.
At the quantum level these symmetries are broken by the 
corresponding anomalies and the anomalous WI are given by 
\bea
\partial^\mu J_\mu &=& - c(g) F^a_{\mu\nu} 
\tilde F^{a\mu\nu},~~
\Theta^\mu_\mu = c(g) F^a_{\mu\nu} F^{a\mu\nu} \no
\gamma^\mu S_\mu &=& 2 c(g) \sigma_{\mu\nu} F^a_{\mu\nu} \lambda^a
\label{anomalies}
\eea
where $J_{\mu}$ and $S_{\mu}$ are the chiral
and SUSY currents respectivley, 
$\Theta_{\mu}^{ \nu}$ the energy momentum tensor, with
$c(g) = \beta(g)/2 g$ and $\beta(g)$ is the beta function of N=1 SYM.
The above two anomalies and the SUSY trace anomaly belong to the 
same supermultiplet.

The corresponding low energy theory of VY \cite{VY}
was obtained by considering the chiral superfield $S$ whose 
components are given by a complex scalar field $\phi$,
a Dirac fermion $\chi$ and complex auxiliary field $M$.
In terms of the fundamental fields, they are described by \cite{VY}
\be
\left . \begin{array}{ccc}
\phi &=& c(g) \bar \lambda^a_R \lambda^a_L \,,\\
\chi &=& \frac{i c(g) }{2} \sigma_{\mu\nu} F^{a\mu\nu} \lambda^a \,,\\
   M &=& - \frac{c(g)}{2} \left ( F^a_{\mu\nu} F^{a\mu\nu} 
                           + i F^a_{\mu\nu} \tilde F^{a\mu\nu} \right )\,, 
\end{array} \right .
\ee
where $c(g)$ is the same factor appearing in the anomalies in 
Eq.(\ref{anomalies}).

The expression of the VY action in terms of the 
superfield $S$ is given by \cite{VY}
\bea
S_{VY} &=& \int d^4 x \left \{ \frac{9}{\alpha} (S^\dagger S)^{1/3}_D 
\right.\no
&+&\left. 
 \left [ \frac{1}{3} \left ( S \log (\frac{S}{\mu^3}) - S \right )_F + h.c. 
\right ] \right \}
\nonumber
\eea
where $\alpha$ and $\mu$ are two free parameters.
Note that the request to reproduce the correct anomalies of
the fundamental action in Eq.(\ref{SYM}) fixes completely the form of the 
superpotential.
The spectrum can be easily analysed by looking at the minimum  
of the scalar potential 
($V_{VY}$) in the exponential representation for 
the scalar field $\phi\equiv \rho e^{i \theta}$
\[
\label{V_VY}
V_{VY} =\frac{\alpha^3}{81} \frac{\rho^4}{4} \left[
\log^2 (\frac{\sqrt{\alpha}}{3\sqrt{2}} \frac{\rho}{ \mu})+
\theta^2 \right].
\]
Then the following conclusions are drawn \cite{VY}
\begin{itemize}
\item
$\mbox{min}(V_{VY})$ is obtained at a non-zero value of $\rho$: 
spontaneous chiral symmetry breaking occurs.
\item The would-be goldstone boson, $\theta$, is not a massless field: 
the anomaly term in the lagrangian explicitly breaks the chiral symmetry,
providing a mass scale for the supermultiplet.
\item 
SUSY is unbroken: mass degeneracy $m_\theta\,=\,m_\rho\,=\,
m_\chi\,=\,\frac{1}{3} \alpha \mu$.
\end{itemize}

Now we explain how to implement the quenching in the N=1 SYM 
theory \cite{qsusy}.
We extend the method proposed by Bernard and Golterman 
\cite{bg} for quenched QCD
to the Majorana fermions in the adjoint representation of the colour group.\\
In order to cancel the fermion determinant, we introduce a ghost
Majorana field $\eta^a$ which has the same quantum numbers as the gluino
$\lambda^a$, but ``wrong'' (bosonic) spin-statistics.
Then the quenched action $S^q_{SYM}$ is given by
\bea
S^q_{SYM} &=& \int d^4 x \left \{ - \frac{1}{4} F^a_{\mu\nu} F^{a\mu\nu}
+  \frac{i}{2} \bar \lambda^a \gamma^\mu D^{ab}_\mu \lambda^b \right.\no
&+&\left. \frac{i}{2} \bar \eta^a ( i \gamma^\mu \gamma_5 ) D^{ab}_\mu \eta^b 
 \right \}
\nonumber
\eea
Note
that (due to the wrong statistic and Majorana nature)
$\bar \eta^a \gamma^\mu D^{ab}_\mu \eta^b = 0$ (up to total derivatives),
in the same way as 
$\bar \lambda^a \gamma^\mu \gamma_5 D^{ab}_\mu \lambda^b = 0$. 
It is important to stress that 
the $S^q_{SYM}$ is no longer supersymmetric, but it acquires
a new $U(1 \mid 1)$ symmetry \cite{qsusy}. 
Note that $S^q_{SYM}$ violates unitarity
due to the ghost $\eta$ field.\footnote{This is a consequence of the fact
that the quenched approximation violates unitarity}

The $U(1 \mid 1)$ group is a $Z_2$ graded Lie group with both bosonic and 
fermionic generators (the supersymmetric algebra itself obeys a $Z_2$ 
graded Lie group) \cite{dewitt}. 
In a more compact form:
\[
\label{qsymaction2}
S^q_{SYM} = \int d^4 x \left \{ - \frac{1}{4} F^a_{\mu\nu} F^{a\mu\nu}
 + i \bar Q^a_R \gamma^\mu D^{ab}_\mu Q^b_R \right \}
\]
where $Q$ is the doublet $Q^a=( \lambda^a , \eta^a )$.
Then $S^q_{SYM}$ is 
invariant under chiral $U(1 \mid 1)$ transformations, defined as follows:
\bea
Q_R & \rightarrow & 
U Q_R  =  \exp{\left\{  i \frac{\alpha_i \sigma^i}{2} \right\}} Q_R \,,\no
Q_L & \rightarrow&  U^\dagger Q_L
\label{transf}
\eea
where $U^\dagger U = I$ and the $\sigma_{i=1,2,3}$ matrices, which 
are the usual Pauli matrices (with $\sigma_0$ the unity matrix),
belong to the algebra of $U(1 \mid 1)$, where $\sigma_0, \sigma_3$ and 
$\sigma_1, \sigma_2$ correspond to the
bosonic and fermionic generators respectively.
The supertrace $Str$ (invariant under $U(1 \mid 1)$) is defined as
\[
Str \left ( \begin{array}{cc} a & b \\ c & d \end{array} \right ) 
= a - d \,,
\]
where, in general, $a,d$ are complex numbers and $b,c$ complex Grassman
numbers.
From the transformations in Eq.(\ref{transf}) we see that 
four currents are associated to the $U(1\mid 1)$ symmetry, which are
$J^i_\mu = \bar Q^a_R \sigma^i \gamma^\mu Q^a_R$ or in components
\cite{qsusy}
\bea
J_\mu^0 &=&\frac{1}{2} ( i \bar \lambda^a \gamma_\mu \gamma_5 \lambda^a + 
\bar \eta^a \gamma_\mu \eta^a ) \no
J_\mu^+ &=& \bar \lambda^a_R \gamma_\mu \eta^a_R,~~~
J_\mu^- = \bar \eta^a_R \gamma_\mu \lambda^a_R \no
J_\mu^3 &=& \frac{1}{2} (i\bar \lambda^a \gamma_\mu \gamma_5 \lambda^a - 
\bar \eta^a \gamma_\mu \eta^a )\,~~
\nonumber
\eea
From the bosonic statistic of the ghost fields $\eta^a$ it follows that 
only $J_\mu^3$ is anomalous. Indeed for the $J^0_\mu$ anomaly the
fermionic-statistic loop 
versus the bosonic one cancels exactly, while for the $J^3_\mu$ case these 
are summed up.
As for the trace anomaly it can be shown that the ghost contribution to 
the trace of the tensor-energy momentum exactly cancel
the contribution of the gluino loop.

In order to generalize the VY effective lagrangian we introduce 
new composite fields which have particular transformation
properties under $U(1\mid 1)$.
In terms of the gluino and ghost fields these  are given by
\bea
\hat{\phi}&\equiv& \sigma^i \hat{\phi}^i,~~~
\hat \phi^i = c(g) \bar Q^a_R \sigma^i Q^a_L,\no
\hat \chi &=& \frac{i c(g) }{2} \sigma_{\mu\nu} F^{a\mu\nu} Q^a
\label{fields}
\eea
with transformation properties
\be
\hat \phi \to  U \hat \phi U,~~~\hat \chi_R \to U \hat \chi_R
\nonumber
\ee
Then we look for the most general low energy effective lagrangian 
$\cal L$ in terms of the fields in Eq.(\ref{fields}). 
This lagrangian can be decomposed as follows \cite{qsusy}
\[
{\cal L} = {\cal L}_{kin} + {\cal L}_{int} + {\cal L}_{anom}
\]
where ${\cal L}_{kin}$, ${\cal L}_{int}$ are 
invariant under chiral $U(1 \mid 1)$ and naive scale transformations. 
The ${\cal L}_{anom}$ is not invariant, but it is 
completely fixed by requiring to
reproduce the anomalies of the quenched fundamental theory. 
Moreover the anomalous $U_{\sigma_3}(1 \mid 1)$ transformations breaks
$U(1 \mid 1)$ as
\[
U(1 \mid 1) \rightarrow Z_{4 N_c} \times SU(1 \mid 1)
\]
We do not give here the expression for the lagrangian 
$\cal L$ that, however, can be found in Ref. \cite{qsusy}.
We only point out that the coefficients of the $U(1\mid 1)$ invariant 
terms in ${\cal L}_{kin}$ and ${\cal L}_{int}$ are fixed 
by requiring that, in the classical unquenched limit ($\eta^a\to 0$),
this lagrangian approaches continuosly to the corresponding one of VY, 
since supersymmetry should be recovered
in this limit. The only term which is non-analytic
in the unquenched limit and is responsible for the mass splitting 
is the anomalous term ${\cal L}_{anom}$, as explained in Ref. \cite{qsusy}.

Now we look at the spectrum in the exponential representation
$\hat{\phi}=\rho\exp{(i \theta^i\sigma_i)}\equiv\rho\hat{\Sigma}$.
Note that, in terms of the original field $\theta$, we have
$\theta_3=\theta-\tilde{\theta}$ and $\theta_0=\theta+\tilde{\theta}$,
where $\tilde{\theta}$ is a pure $\eta$ ghost condensate and it goes 
to zero in the unquenched limit.

The mass spectrum in terms of the original VY fields 
is obtained by using the technique explained in Ref. \cite{bg} and we find
\be
m_\rho = \frac{\beta^\prime}{\beta} m_\chi,~~~
m_\theta =  (1+1) m_\chi
\label{qspectrum}
\ee
where $\beta^{\prime}(g)$ is
the one-loop $\beta$ function of the pure Yang-Mills theory.
This spectrum should be 
compared with $m_\chi \,=\, m_\sigma \,=\, m_\theta$ in the unquenched theory. 
Note that the splitting in the mass spectrum of Eq.(\ref{qspectrum})
provides an estimation of the error induced by the quenched approximation.
Now we summarize the main conclusion of this analysis \cite{qsusy}
\begin{itemize}
\item
The mass splitting of the VY supermultiplet 
results from the non-analiticity of the anomaly structure 
induced by the ghost field.
\item
The numerical result obtained in Ref.\cite{andrea} 
for the ratio $(m_\rho/m_\chi)_{lat} = 1.1(3)$
is in fair agreement with our theoretical expectation
$(m_\rho/m_\chi)_{th} = 11/9 = 1.22$ for $SU(2)$. 
\end{itemize}
\section{ Strong coupling limit}
The lattice strong coupling expansion is a very powerful 
analytical probe in order to study the critical behaviours of 
lattice gauge theories and also to test qualitatively their 
continuum properties.
The strong coupling expansion technique has been extensively used in
pure Yang-Mills theory and in QCD, often
combined also with the large $N_c$ expansion.
We recently investigated the strong coupling limit at large $N_c$ of a N=1 SYM
theory and now we present the main results and conclusions of this
work \cite{our}.

The usual computational frameworks of strong coupling expansion can 
be summarized as follows
\begin{itemize}
\item 
Effective actions \cite{BreKlub}-\cite{KawSmit}:
The Wilson-Dirac lattice action is considered
at large $N_c$ and small $\beta=1/g_0^2$.
The large $N_c$ expansion can be recognized as a 
saddle-point expansion of the gauge functional integral, 
previously simplified by the $\beta \to 0$ limit. 
\item 
Path resummation \cite{Kawam}-\cite{combi}:
The fermion matrix $\mathbf{M}$ is inverted by using the standard hopping 
parameter expansion, 
which expresses the propagator $\mathbf{M^{-1}}$ 
as a sum over paths on the lattice.
\end{itemize}
Our technique is based on the path resummation formulas
which are valid irrespective of the
representation in which the matter lies and for general $r$.
Moreover we keep $r$ arbitrary since it allows the possibility of 
searching for multicritical points: indeed more freedom in the parameter 
space is necessary in order to search for the simultaneous restoration 
of supersymmetry and chiral symmetry.
Now we present the general formalism.

The SUSY Yang-Mills action on the lattice can be formally written as
\[
S = \beta S_g + \frac{1}{2} \Psi_i \Psi_j M_{i j }\ \ ,  
\]
where $\beta S_g$ is the pure gauge part and $\Psi_i$ is a Grassman variable
representing  the field of a Majorana fermion. The matrix $\tilde{M}$ must be
antisymmetric and its form is given by 
\bea
\tilde{M} &=& C M,~~~M = \mathbf{I} - \sum_{\alpha \in I} 
\mathbf{\Delta}_{\alpha}\no
(\mathbf{\Delta}_{\alpha})_{i j }&=& k\delta_{m\, n+V(\alpha)}\
U_{\alpha}^{a b}(n)\,(r \mathbf{I} - \gamma_{\alpha})_{A B}
\nonumber
\eea
where $U$ is the gauge link variable,
$\mathbf{I}$ and $C$ are the unity and charge conjugation matrix respectively
and $\kappa$ is the hopping parameter.
With the indices $i$ and $j$ we simbolically indicate
$i=(n,a,A)$, where $n,~a$ and $A$ 
run on the lattice points, the indices of the $SU(N_c)$ 
adjoint representation, and the Dirac indices respectively. 
Now we will concentrate upon the gauge invariant operators
of the form:
\be
\bO_i(x)=\Psi_{A_1}^{a_1}(x) \ldots \Psi_{A_p}^{a_p}(x)\,
(\cS_i)_{A_1 \ldots A_p}\, \cC_i^{a_1 \ldots a_p} 
\label{pgluino}
\ee
where $\cC_i^{a_1 \ldots a_p}$ is an invariant color tensor and
$(\cS_i)_{A_1 \ldots A_p}$ a spin tensor. For $p=2$ a basis
for $S_i$ is the Clifford algebra basis in d dimension.
\\
We are interested in computing the following quantities at strong coupling
\be
\langle\bO_i(x)\rangle,~~~~
G_{i j }(x-y) \equiv \langle\bO_i(x) \bO_j(y)\rangle
\nonumber
\ee
where as usual the $\langle~\rangle$ means the vacuum expectation value.

We will be able to accomplish this goal  for $\beta=0$
and in the large $\NC$ limit, and the
corrections to the formulas in powers of 
$\beta$ and $\frac{1}{\NC}$ are  in principle feasible.
When the fermion are integrated out we obtain
\bea
&& \prod_{i}\left( \int  d\Psi_i\right)\, \exp\{-\frac{1}{2} \Psi_i \Psi_j\,
\bM_{i j } + J_i \Psi_i \} =\no
&&Pf(\bM) \exp\{-\frac{1}{2} J_i  J_j
(\cM^{-1} C^{-1})_{i j} \}
\eea
where $Pf(\bM)$ stands for the Pfaffian of the matrix $\bM$.
\footnote{
The square of the Pfaffian is the determinant, and up to a sign
\[
Pf(\bM)= \sqrt{\det(C) \det(\cM)} = \exp\{\frac{1}{2}
Tr(\log(\cM)) \} 
\]
We checked that $Pf(\bM)$ is always positive
provided that $|\kappa| < \frac{1}{2 d\, (|r|+1)}$}
Now the next step is to expand the previous quantities as a sum over
all the possible paths $\gamma$ going from $x$ to $y$
\bea
&&(\cM^{-1}(x,y))^{a b}_{A B}=\sum_{\gamma \in \cS(x \rightarrow
y)} W^{a b}(\gamma)\,
\Gamma_{A B}(\gamma)\no
&&Pf(\bM)=\exp\{\frac{1}{2}\sum_{x \in {\cal L}} \sum_{L=1}^{\infty}
\sum_{\gamma \in \cS_L(x \rightarrow x)} \no
&&\frac{1}{L}\, Tr(W(\gamma))\, Tr(\Gamma(\gamma))\}
\nonumber
\eea
where $ W(\gamma)$ is the path ordered product (along the path $\gamma$) of the gauge field link variables $U(x)_{\alpha}^{ab}$ and
$\Gamma(\gamma)$ denotes the appropriate product of the spin matrices:
\[
\Gamma(\gamma\equiv (x,\vec{\alpha}))= \kappa^L (r -\gamma_{\alpha_1})
\cdots  (r -\gamma_{\alpha_L})
\]
where $L$ is the lengh of the path in lattice space unity and
$S_L(x\to y) $ in the sum indicates the sum over the paths of the lenght L
which goes from $x$ to $y$.
Now we present the main results obtained by using the results 
of Ref.\cite{largeN} for the $SU(\NC)$ group integration for 
gauge fields in the adjoint representation at large $\NC$~:
\begin{itemize}
\item 
The quenched approximation and the OZI approximations
turn out to be exact in the large $N_c$ limit.
\item 
The results obtained at $\beta_{adjoint}=0$
are exact in the large $N_c$ limit: 
the corrections of $O(\beta)$ are subleading in the large $N_c$ limit.
\end{itemize}
Now we give the main formulas (at large $\NC$) 
for the condensates $\langle O_i(x)\rangle$ and
propagators of the 2-gluino operators
$G_{i j}(x)$, defined in Eq.(\ref{pgluino}) ( with $p=2$), obtained
after resumming over paths:
\bea
&& \langle {\bf O_i}(x) \rangle~=~R_1(\xi) Tr(\hat{S}_i)\no
&&G_{i j}(x) =R_2(\xi)\,
\prod_{\mu} (\int \frac{d\varphi_{\mu}}{2 \pi})\, e^{\imath \varphi \cdot x}
\no
&&
\times 
\langle \cS_i |\,\CTT^{-1} \, \lbrack \Theta_2(\xi)\mathbf{I} - \tAtwo(\varphi)
\rbrack^{-1} | \cS_j \rangle\no
 \tAtwo(\varphi) &\equiv& \kappa^2\, \sum_{\alpha \in I}  e^{\imath
\varphi_{\alpha} }
 \rmga \otimes \rmga 
\label{propagator}
\eea
where  $\CTT \, \equiv \, C^{-1} \otimes C^{-1}$ with $C$ 
the charge conjugation matrix, 
$\xi$ is a function of $r,k$ given in Ref.\cite{our}, and
$S_i$ are matrices of the Clifford algebra basis in d-dimension.
The expressions for the functions $R_2(x)$ and $\Theta_2(x)$ can be found 
in Ref.\cite{our}. 
We have analogous expressions for the p-gluino propagators 
provided that the function $R_2$, the 
vectors $|S_i\rangle$ and the matrix $\tAtwo(\varphi)$,
appearing in Eq.(\ref{propagator}), are substituted with the
 corresponding ones for the p-gluino operators.

The main difficulty in order to calculate the propagators 
is given by the 
calculation of the inverse matrix $\Theta_p(\xi)\mathbf{I}-\tAp(\varphi)$
for general p-gluino operators.
This goal has been achieved for the 2-gluino operators
in any dimension by means of the gamma-fermions techniques developed 
in Ref.\cite{our}.
In general for the p-gluino sector (with $p>2$) we have been able to invert 
this matrix only in the particular limit where the spectrum 
of the p-gluino propagators becomes degenerate.

In general the procedure to obtain the masses can be summarized as follows:
extract the eigenvalues of the matrix
$\Theta_p(\xi)\mathbf{I}-\tAp(\vec{\varphi}=\vec{0})$, 
which are functions of the temporal momentum $\varphi_0$.
Then determine $\varphi_0^{pole}$ which is 
the (complex) value of $\varphi_0$ for
which the eigenvalues vanishes.  Finally the lattice masses are given by
\be
M=-log(|e^{\imath \varphi_0^{pole}}|).
\ee
Note that the lattice masses 
are dimensionless quantities and depend only on $k$ and $r$.
The physical masses are proportional to $M/a$ and so
the states whose lattice mass vanish at the critical line, are
the states that survive this continuum limit.

Now we present below the main results for the spectrum of the 2-gluino and
3-gluino operators in d=4.
\begin{itemize}
\item 
Chiral symmetry is spontaneusly broken.
\item 
The pseudoscalar is the lightest states and the critical lines 
where the scalar or the lightest fermion become massless are outside of the
physical region in the $(k,r)$ plane.
\item 
All the meson masses become degenerate only for
$r\to \infty$ and $\kappa \to 0$ with the product $\kappa r =fixed$.
In particular all the mesons become massless in the limit
where $\kappa r=\frac{1}{2\sqrt{2d-1}}$.
\item
In this limit~: the lightest fermion mass can not be degenerate with 
the lightest meson sector and for $p>q$ any mass in the $p$-gluino sector 
(the fermions have $p$ odd) is 
higher than any other in the $q$-gluino sector in this limit.
\end{itemize}
From these results we argue that there are no points 
in the $(k,r)$ plane giving a possible candidate for a supersymmetric continuum
limit.
\section{Conclusions}
In order to estimate the error induced by the 
quenched simulations on the spectrum, 
we implemented the quenching in the fundamental theory by introducing a 
ghost field. 
Although SUSY is lost upon quenching, it turns out that a new 
$U(1\mid 1)$ symmetry arises, explicitly broken by the chiral
anomaly to $Z_{4N_c}\times SU(1\mid 1)$. 
Then we carried out this new symmetry and the corresponding anomalies 
in a low energy lagrangian scheme.
The result is that the anomaly structure entails a 
controlable splitting of the VY multiplet by giving the scalar mass 
$20\%$ heavier then the fermionic one. These results
are in fair agreement with the  numerical quenched ones within the 
statistical errors and provides
a first estimate of the systematic error associated to the
quenching in lattice SUSY computations.

From the side of the strong coupling limit at the large $N_c$~:
we used the hopping parameter ($k$) expansion in terms of random walks
resummed for any value of the Wilson parameter $r$ and 
close to the origin in $k$.
We found exact analytical results for the condensates, 
propagators and spectrum in the large $N_c$ limit,
for arbitrary dimensions and general $r$.
By analysing these results our main conclusion 
is that the quenched and the usual 
OZI approximations are exact at $\beta_{adjoint}=0$ 
and in the large $N_c$ limit.
Moreover we proved that in the strong coupling
regime there are no critical lines or points in the
$(k,r)$ plane giving a possible candidate for a supersymmetric continuum
limit, at least in the validity region of the hopping parameter expansion
close to the origin.
\section*{Acknowledgments}
I thank I. Antoniadis, A. Masiero and S. Pastor
for the invitation to this meeting. 
I thank also A. Donini, B. Gavela, M. Golterman,
A. Gonz\'alez-Arroyo and C. Pena for useful
discussions. 
I acknowledge the financial support of the TMR network project 
ref. FMRX-CT96-0090.

\end{document}